\documentclass[aps,prb,twocolumn]{revtex4}

\usepackage{graphicx}
\def  \bsig    {\mbox{\boldmath$\sigma$}}

\begin{document}

\title{Equilibrium spin currents and magnetoelectric effect
in magnetic nanostructures}
\author{Patrick Bruno$^1$ and Vitalii K.~Dugaev$^{2,3}$}
\affiliation{$^1$Max-Planck-Institut f\"ur Mikrostrukturphysik,
Weinberg 2, 06120 Halle, Germany\\
$^2$Department of Physics and CFIF, Instituto Superior T\'ecnico,
Av. Rovisco Pais, 1049-001 Lisbon, Portugal\\
$^3$Frantsevich Institute for Problems of Materials Science,
National Academy of Sciences of Ukraine, Vilde 5, 58001
Chernovtsy, Ukraine}
\date{26 August 2005}

\begin{abstract}
We discuss the problem of equilibrium spin currents in
ferromagnets with inhomogeneous magnetization. Using simple
microscopic models we explain the physical origin of equilibrium
spin currents. Next we derive the equilibrium spin current from
the Hamiltonian with a gauge field associated with local rotations
in the spin space. Several examples of magnetic systems are
studied in details, and the persistent spin current is found to
exist in the ground state of these systems. We also demonstrate
the possibility to measure the equilibrium spin current using the
magnetoelectrically induced electric field near the ring.

\vskip0.5cm \noindent
\end{abstract}
\maketitle

1. {\it Introduction}. The problem of generation of the pure spin
currents attracted a lot of attention recently. It is mostly
related to the perspectives of possible applications in
spintronics, for which the generation and manipulation of spin
currents is of a prime importance.\cite{prinz98} Several different
ways have been proposed to create the spin currents, like, for
example, the injection of spin-polarized carriers from a magnetic
metal or semiconductor,\cite{ohno99} optical excitation of
spin-polarized electrons or holes,\cite{ganichev02} equilibrium
spin currents induced by the spiral states in
ferromagnets,\cite{konig01} spin Hall effect,\cite{hirsch99} spin
transport in presence of the spin-orbit (SO)
interaction,\cite{pareek04} and others.

Recently, the transport of magnetization by magnons has been
studied by Meier {\it et al}.\cite{meier03} They demonstrated that
by using a finite-length spin chain between magnetic reservoirs,
the pure spin current can be generated without the transport of
electron charge. Then Sch\"utz {\it et al} proposed to use the
magnons in a mesoscopic Heisenberg ring under an inhomogeneous
magnetic field as a possible way to induce the persistent spin
current.\cite{schutz03} It was also found in these
works\cite{meier03,schutz03} that the magnetization transport by
magnons is accompanied by an electric polarization, which can be
experimentally observed in the vicinity of the magnetic
wire\cite{meier03} or the mesoscopic ring.\cite{schutz03} Most
recently, Katsura {\it et al} used an electronic model of two
transition atoms mediated by an oxygen atom, to show that the
equilibrium spin current in this system is related to the electric
polarization through a new mechanism of the magnetoelectric
effect,\cite{katsura05} which, in fact, is much stronger than the
previously discussed\cite{meier03,schutz03} relativistic
mechanism.

At the same time, the theoretical studies of the magnetization
transport provoked a lot of discussions about the existence and
physical meaning of the non-vanishing equilibrium spin currents
and about a proper definition of the spin current in the
non-equilibrium transport
phenomena.\cite{rashba03,schutz04,zhang05} The main idea of these
works was to re-define the transport spin current, which does not
include the equilibrium (supercurrent) part responsible for the
spin torque.\cite{zhang05}

Here we show that several key points concerning the equilibrium
spin currents and electric polarization discussed in the
above-mentioned papers\cite{meier03,schutz03,schutz04,katsura05}
should be further clarified. In particular, we emphasize that the
equilibrium spin current (also called supercurrent because it is
associated with the rigidity of the order
parameter\cite{chandra90}) is related to a noncollinear magnetic
ordering, and it exists at $T=0$ without magnons (in contrast to
the claims of Refs.~[\onlinecite{schutz03,schutz04}]). The details
of electronic structure and/or SO interaction are not important
for the equilibrium spin currents. On the other hand, the
observable electric polarization is induced by the equilibrium
spin current as well as by the magnon transport, and the
corresponding mechanism necessarily includes the spin-orbit
interaction. The magnitude of this effect strongly depends on the
electronic structure of the material, so that it can be much
larger than a bare relativistic effect.\cite{katsura05} We believe
that the effective enhancement of the SO coupling for the
magnetoelectric mechanism in the noncollinear magnets is an
analogue of the renormalization (giant enhancement) of the Rashba
SO interaction in the layered systems, as well as the
reinforcement of the anomalous Hall effect, which is also related
to the SO interaction.

2. {\it Two spins in local fields}.
We start from a simple example demonstrating that the equilibrium
spin current does not vanish
for two interacting quantum spins $S=1/2$.
Namely, we consider the Hamiltonian with interacting spins $S_1$
and $S_2$ in local magnetic fields ${\bf B}_1$ and ${\bf B}_2$
\begin{equation}
\label{1}
H=-J\, {\bf S}_1\cdot {\bf S}_2
-{\bf B}_1\cdot {\bf S}_1
-{\bf B}_2\cdot {\bf S}_2\, ,
\end{equation}
where $\left| {\bf B}_1\right| = \left| {\bf B}_2\right| =B$.
Calculating the time-derivative of spin operator
$\dot{S}_{1\mu}=\frac{i}{\hbar }\; \left[ H,\, S_{1\mu }\right] $
with (1), we find
\begin{equation}
\label{2}
\dot{S}_{1\mu}
=\frac{J}{\hbar }\; \epsilon _{\mu\nu\lambda}\;
S_{1\nu}\; S_{2\lambda }
+\, \frac1{\hbar }\; \epsilon _{\mu\nu\lambda}\;
S_{1\nu}\; B_{1\lambda }\, .
\end{equation}
The first term in the right side of (2) is the variation
of spin related to the spin current $j^\mu _{1\rightarrow 2}$,
and the second one is related to the local field ${\bf B}_1$
acting on ${\bf S}_1$.
In the equilibrium
state, $\left< \dot{S}_{1\mu}\right> =0$ due to compensation
of the moments transferred by the spin current and generated by the
field ${\bf B}_1$.

To calculate the value of spin current in the ground state, we use
the perturbation theory assuming the interaction weak, $\left|
J\right| /B\ll 1$. Let us take the quantization axis $z$
along the field ${\bf B}_1$, and the axis $y$ in the plane of
vectors ${\bf B}_1$ and ${\bf B}_2$. In the limit of $\left|
J\right| /B\ll 1$, the eigenfunction of Hamiltonian (1)
corresponding to the ground state with energy $\varepsilon
=-2B$ is $\left< \psi \right| =\left< \psi _1\right|
\otimes \left< \psi _2\right| $ with
\begin{eqnarray}
\label{3}
\left< \psi _1\right|
=\left( 1,\, 0\right) ,\hskip0.2cm
\left< \psi _2\right|
=\left( \cos \frac{\theta }2,\, -i\sin \frac{\theta }2\right) ,\hskip0.3cm
\end{eqnarray}
where $\theta $ is the angle between the vectors ${\bf B}_1$ and
${\bf B}_2$.

Calculating the spin current in this state we find
$\left< j^x_{1\rightarrow 2}\right>
=-J\sin \theta /\hbar $,
and
$\left< j^y_{1\rightarrow 2}\right> =\left< j^z_{1\rightarrow 2}\right> =0$.
The similar calculation gives us $\left< j^x_{2\rightarrow 1}\right> =
-\left< j^x_{1\rightarrow 2}\right> $, which means nonconservation
of the spin current in the presence of the local fields ${\bf B}_1$
and ${\bf B}_2$. The spin current transferred to the spin ${\bf S}_1$
generates the torque rotating this spin from the orientation along
vector ${\bf B}_1$.

3. {\it Hubbard model}. This result can be explained by
microscopic calculation using a two-site electronic model with
Coulomb interaction. Let us consider a model of two-electron
system with strong on-site Hubbard interaction $U$. This is an
example of an electronic system, which is often used as a model
leading to the localized magnetic moments with antiferromagnetic
interaction. We show that by applying on-site magnetic fields
${\bf B}_1$ and ${\bf B}_2$ with different orientations, we induce
the spin current transferring a spin moment. The Hamiltonian is
\begin{eqnarray}
\label{4}
H=-t\left( c^\dag _{1\alpha }\, c_{2\alpha }
+c^\dag _{2\alpha }\, c_{1\alpha }\right)
+U\, \sum _{i=1,2} \left( c^\dag _{i\uparrow }\, c_{i\uparrow }\,
c^\dag _{i\downarrow }\, c_{i\downarrow }
\right. \nonumber \\ \left.
-B^\mu _i\,
c^\dag _{i\alpha }\, \sigma ^\mu _{\alpha\beta }\, c_{i\beta }\right) ,
\hskip0.5cm
\end{eqnarray}
where $t$ is the hopping, $c^\dag _{i\alpha }$ and
$c_{i\alpha }$ are the creation and annihilation operators for
electrons at the site $i$ with spin $\alpha $.
We take $\left| {\bf B}_1\right| =\left| {\bf B}_1\right| =B$ and
consider the case of strong on-site interaction and weak field,
$B\ll t\ll U$.

The operator of the spin at site 1 is ${\bf S}_1=\frac12 c^\dag
_{1\alpha }\bsig _{\alpha\beta}c_{1\beta }$, and the spin dynamics
is determined by $\dot{\bf S}_1=i\left[ H,\, {\bf S}_1\right] $,
where the dot means time derivative. Using (4) we find
\begin{eqnarray}
\label{5}
\dot{S}^\mu _1
=\frac{it}2 \left( c^\dag _{1\alpha}\, c_{2\beta}
-c^\dag _{2\alpha}\, c_{1\beta}\right) \sigma ^\mu _{\alpha\beta}
+\epsilon ^{\mu\nu\lambda} \sigma ^\nu _{\alpha\beta}
B^\lambda _1 c^\dag _{1\alpha} c_{1\beta}\, .\hskip0.3cm
\end{eqnarray}
Here the first term in the right hand site is the operator of spin
current transferring spin from site 2 to site 1. The second term
is related to the force acting on the spin ${\bf S}_1$ by the the
field ${\bf B}_1$. In the ground state $\left< \dot{\bf
S}_1\right> =0$, which can be satisfied due to the nonvanishing
spin current
\begin{equation}
\label{6}
j^\mu =\frac{it}2 \, \sigma ^\mu _{\alpha\beta}
\left< c^\dag _{1\alpha}\, c_{2\beta}
-c^\dag _{2\alpha}\, c_{1\beta}\right> _0 ,
\end{equation}
flowing from site 2 to site 1. The spin current transfers the
angular moment compensating the force produced by the field
${\bf B}_1$.

We consider the system with two electrons, and take the basis of
two-particle states $\left| \uparrow \downarrow ,0\right> $,
$\left| 0,\uparrow \downarrow \right> $, $\left| \uparrow
,\uparrow\right> $, $\left| \uparrow ,\downarrow\right> $, $\left|
\downarrow ,\uparrow\right> $, $\left| \downarrow
,\downarrow\right> $. The Hamiltonian (4), and the operators of
spin ${\bf S}_i$ and the spin current ${\bf j}$ can be presented as
matrices in the basis of two-particle states. For example, if
we take ${\bf B}_1$ along $z$, and ${\bf B}_2$ in the $y-z$ plane,
the Hamiltonian in the new basis acquires the following form
\begin{equation}
\label{7}
H=\left( \begin{array}{cccccc}
U & 0 & 0 & t & -t & 0\\
0 & U & 0 & -t & t & 0\\
0 & 0 & -B-B_z^\prime & iB_y^\prime & 0 & 0\\
t & -t & -iB_y^\prime & -B+B_z^\prime & 0 & 0\\
-t & t & 0 & 0 & B-B_z^\prime & iB_y^\prime \\
0 & 0 & 0 & 0 & -iB_y^\prime & B+B_z^\prime
\end{array}
\right) ,
\end{equation}
where we denoted $B=B_{1z}$ and $B_\mu ^\prime =B_{2\mu }$.

In the limit of strong interaction, $t/U\ll 1$, and
$B=0$, this Hamiltonian can be transformed by a unitary transformation
to the block form, in which the part describing  the spin
coupling is $\tilde{H}=J\, ({\bf S}_1\cdot {\bf S}_2-1/4)
-{\bf B}_1\cdot {\bf S}_1-{\bf B}_2\cdot {\bf S}_2$, where $J=4t^2/U$.
By using the same transformation to the spin-current matrix,
we find that it transforms into $\tilde{\bf j}=J\, {\bf S}_1\times {\bf S}_2$.
This calculation shows that, microscopically,
the equilibrium spin current results
from the hopping of electrons between different sites, with an
effective transfer of moment by the spin of electrons.

4. {\it Noncollinear ferromagnet}. Now we consider a textured
ferromagnet, which represents a magnetic system in a topologically
nontrivial metastable state. A simple example of such system is a
magnetic ring with the easy-axis magnetic anisotropy, so that the
magnetic moment, oriented along the ring, creates a vortex.

Let us take a continuous classical model of a ferromagnet
described by the Hamiltonian, which includes the exchange
interaction, anisotropy, and the interaction with an external
magnetic field ${\bf B}({\bf r})$ (this model was used recently to
determine the Berry phase of magnons in textured
ferromagnets\cite{bruno04})
\begin{eqnarray}
\label{8}
H=\int d^3{\bf r}\left[ \frac{a}2\left(
\partial _i\, n_\mu \right ) ^2 +\mathcal{F}\left\{ {\bf n({\bf
r})}\right\}
-\, \beta \, B_\mu n_\mu\right] ,\hskip0.3cm
\end{eqnarray}
where ${\bf n}({\bf r})$ is the unit vector oriented along the
magnetization at the point ${\bf r}$, $a$ is the constant of
exchange interaction, $\mathcal{F}\left\{ {\bf n}({\bf
r})\right\}$ is a function determining the anisotropy.

We will use the definition of spin current related to the
transformation of the Hamiltonian under local rotations of vector
${\bf n}$ in the spin space.\cite{chandra90} This definition is in
accordance with a general definition of currents in the quantum
field theory.\cite{itzykson} In the case of a classical magnetic
system with Hamiltonian (8), the corresponding transformations of
the vector ${\bf n}({\bf r})$ belong to the group SO(3). Thus, we
perform a local rotation ${\bf n}({\bf r})\rightarrow {\rm R}({\bf
r})\, {\bf n}({\bf r})$ using the orthogonal transformation matrix
${\rm R}({\bf r})=e^{i \psi ({\bf r})J^z} e^{i \theta ({\bf
r})J^y} e^{i \phi ({\bf r})J^z}$, where $\psi $, $\theta $, $\phi
$ are the Euler angles determining the arbitrary rotations of the
coordinate frame, and $J^x$, $J^y$, and $J^z$ are the generators
of 3D rotations around $x$, $y$ and $z$ axes, respectively.

The Hamiltonian of exchange interaction (the first term in
Eq.~(8)) in the rotated frame has the form
\begin{equation}
\label{9}
H_{ex}=\frac{a}2 \int d^3{\bf r} \left( \partial _i n_\mu
-A_{i\, {\mu\nu}}\, n_\nu \right) ^2,
\end{equation}
where the gauge field
$A_i({\bf r})=\left( \partial _i\, {\rm R}\right) {\rm R}^{-1}$.
The matrix $A_i({\bf r})$ can be presented as
$A_i({\bf r})=iJ^\mu \mathcal{A}^\mu _i({\bf r})$,
where $\mathcal{A}^\mu _i({\bf r})$ belongs to the adjoint
representation of the group SO(3).
Then the exchange energy can be written as
\begin{equation}
\label{10}
H_{ex}=\frac{a}2 \int d^3{\bf r} \left[ \left(
\partial _i\; \delta _{\alpha\beta}
-i\mathcal{A}^\mu _i\, J^\mu _{\alpha\beta}\right) n_\beta \right]
^2,
\end{equation}
and the spin current density is defined as
$j^\mu _i=\gamma \, (\delta H/\delta \mathcal{A}^\mu _i)$,
where $\gamma $ is the gyromagnetic ratio. We find
\begin{equation}
\label{11}
j^\mu _i=-ic_sJ_{\alpha\beta}^\mu n_\beta \left(
\partial _i\; \delta _{\alpha\gamma}
-i\mathcal{A}^\delta _i\, J^\delta _{\alpha\gamma}\right) n_\gamma
\, ,
\end{equation}
where $c_s=\gamma a$. The spin current (11) is gauge invariant. We
can fix the gauge by taking the auxiliary field $\mathcal{A}^\mu
_i=0$. Then, using the relation $iJ^\mu _{\nu\lambda}=\epsilon
_{\mu\nu\lambda}$, we finally obtain
\begin{equation}
\label{12}
{\bf j}_i=c_s\, {\bf n}\times \partial _i {\bf n}.
\end{equation}
Thus, the spin current (12) is nonzero in the noncollinear
ferromagnets. In particular, it is nonzero in a metastable state
of the ferromagnet with topological excitations.\cite{pokrovsky88}
This definition is consistent with the spin-conservation equation
relating the variation of the magnetization in time to the
divergence of the spin current,\cite{durst00,schutz03} $\dot{n_\mu
} +{\rm div}\, {\bf j}^\mu =T_{\mu }$, where $T_\mu =\gamma
\epsilon _{\mu\nu\lambda}\, n_\nu B_\lambda ^{eff}$, and ${\bf
B}^{eff}$ is the local effective magnetic field, comprising the
local external field and other local terms, such as magnetic
anisotropies. The physical meaning is that of a hydrodynamic
equation for ${\bf n}({\bf r})$, where $T_\mu $ is a local source
term (torque).

The thermodynamic average of the spin current is
\begin{equation}
\label{13}
\left< j_i^\mu \right>
=c_s\epsilon _{\mu\nu\lambda}\left[
\left< n_\nu \right>
\partial _i\left< n_\lambda \right>
+\left< \delta n_\nu \, \partial _i\, \delta n_\lambda \right> \right]
\end{equation}
where $\left< X\right> \equiv {\rm Tr}\, (Xe^{-\beta H})/{\rm
Tr}\, (e^{-\beta H})$, and $\delta X=X-\left< X\right> $ is the
fluctuation of $X$. Thus, there are two contributions to the spin
current: the principal one arises from spatial variations of the
average value of the magnetization axis, while the second one is
due to the magnetization fluctuations (magnons). The magnon term
is likely to be a small correction as compared to the principal
(ground state) term. The ground state term can be interpreted as
the supercurrent associated with the rigidity of the order
parameter.\cite{chandra90,konig01,tatara03} The magnon term has been
discussed thoroughly,\cite{meier03,schutz03,wang04} and will not
be discussed further here. In the following, we assume $T=0$, so
that the magnon term vanishes (except for the effect of quantum
zero-point fluctuations, which are usually negligibly small in
ferromagnets). In this case, we can drop the angular brackets,
which means that the thermodynamic average is implied.

5. {\it Mesoscopic ring}.
To calculate the spin current in the ring geometry, it is
convenient to use the cylindric
coordinates $(\rho ,\, \varphi ,\, z)$ for the point on the ring.
We assume that the vector ${\bf n}$ does not depend on $\rho $ and
$z$. Then the exchange Hamiltonian $H_{ex}$ can be written as
\begin{equation}
\label{14}
H_{ex}=\frac{a\, \zeta _0}{2R} \int _0^{2\pi }d\varphi
\left( \partial _\varphi n_\mu +\frac1{2\pi }\;
\epsilon _{\mu\lambda\nu}\, \Phi _\lambda \, n_\nu \right) ^2,
\end{equation}
where $R$ and $\zeta _0$ are the radius and the cross section of
the ring, respectively, $\Phi _\lambda \equiv L\mathcal{A}^\lambda
_\varphi $, and $L=2\pi R$.

Let us take $\mathcal{A}^\mu _\varphi $ constant along the ring.
Then $\Phi _\mu $ is the flux of the $\mu $-component of the gauge
field $\mathcal{A}^\mu _\varphi $ through the ring. Now we get the
azimuthal spin current density in the ring as $j^\mu _\varphi
=(\gamma /\zeta _0)\, (\partial H/\partial \Phi _\mu )$. Using
(14) and taking $\Phi _\mu =0$, we find
\begin{eqnarray}
\label{15-17}
j^\rho _\varphi =\frac{c_s}{L}\int d\varphi \left[
-n_z\left( n_\rho +\partial _\varphi n_\varphi \right)
+n_\varphi \; \partial _\varphi n_z
\right] ,\hskip0.5cm
\\
j^\varphi _\varphi =\frac{c_s}{L}\int d\varphi \left[
n_z\left( -n_\varphi +\partial _\varphi n_\rho \right)
-n_\rho \; \partial _\varphi n_z \right] ,\hskip0.5cm
\\
j^z _\varphi =\frac{c_s}{L}\int d\varphi \left[
n_\varphi \left( n_\varphi -\partial _\varphi n_\rho \right)
+n_\rho \left( n_\rho+\partial _\varphi n_\varphi \right)
\right] .\hskip0.3cm
\end{eqnarray}

In particular cases of the tangential magnetization
($n_\varphi =1$, $n_\rho =n_z=0$), radial magnetization
($n_\rho =1$, $n_\varphi =n_z=0$), and also for any intermediate
case with the in-plane magnetization making a constant
angle with the tangent vector, using (15)-(17) we obtain
\begin{equation}
\label{18}
j^z_\varphi =2\pi c_s/L\, ,\hskip0.5cm
j^\rho _\varphi =j^\varphi _\varphi =0.
\end{equation}
It should be emphasized that the spin current (18) is related to
the assumed metastable state of magnetization field
but not to the magnons. The contribution of magnons
exists for $T\ne 0$ but it is small because the magnons are weak
excitations over the metastable state, completely vanishing in the
limit of $T\rightarrow 0$.

The other example is a magnetic ring with uniaxial anisotropy in a
homogeneous magnetic field along the axis $z$. Due to the
anisotropy and exchange interaction, the magnetization along the
ring is oriented with a certain angle out of the ring plane.
There exist a metastable state a with crown-like magnetization
profile in this
system. (Due to the dipolar forces it can be the ground state.)
We can calculate the angle $\theta $ using Hamiltonian (8) with
$\mathcal{F}\{{\bf n}({\bf r})\}=\lambda \, n_z^2/2$ and
$\lambda >0$, and the
magnetic field ${\bf B}$ oriented along the axis $z$. Then,
using the polar coordinates and assuming that $n_\varphi $ and
$n_z$ do not depend on the coordinate along the ring, we find the
energy of the metastable state
\begin{equation}
\label{19}
E=\pi \zeta _0R\left[
an_\varphi ^2/(2R^2)+\lambda n_z^2/2-\beta Bn_z\right]
\end{equation}
and calculate the angle $\theta $ minimizing the energy (19),\cite{bruno04}
$\cos \theta ={\rm max}\, \left\{1,\; \beta B/(\lambda -a/R^2)\right\}$
for $\lambda >a/R^2$, and $\theta =0$ for $\lambda \leq a/R^2$.
The spin current in the ground state with $\theta \ne 0$ does not
vanish. We calculate the components of it using Eqs.~(15)-(17)
\begin{equation}
\label{20}
j^\varphi _\varphi =\frac{2\pi c_s}{L}\; \sin \theta \, \cos \theta ,\hskip0.2cm
j^z _\varphi =\frac{2\pi c_s}{L}\; \sin ^2\theta ,\hskip0.2cm
j^\rho _\varphi =0.
\end{equation}
Obviously, we obtain the same value of the spin current for different
magnetization profiles if they can be transformed to
each other by a global rotation.

6. {\it Electric polarization}. As discussed in
Refs.~[\onlinecite{meier03,schutz03}], the spin current implies an
electric polarization $P_i=\epsilon _{ij\mu } j_j^\mu /c^2$, which
is the relativistic effect of a transformation of magnetization to
the electric field in the moving frame (this fact that has been
known long ago\cite{frenkel26}). The important point is that the
electric polarization appears not only due to the spin transport
of magnons but also due the spin supercurrent in the ground state
(first term in Eq.~(13)). This was shown by Katsura {\it et
al.}\cite{katsura05} using a microscopic model with two transition
atoms and an oxygen atom in-between. Their model calculations
demonstrate that the magnetoelectric coefficient determining the
magnitude of effect depends essentially on the material
parameters. Taking it into account, we can write the polarization
\begin{equation}
\label{21} {\bf P}=\alpha _{me}\left[ {\bf n} \left( \nabla \cdot
{\bf n} \right) -( {\bf n} \cdot \nabla )\,  {\bf n} \right] ,
\end{equation}
where we denoted by $\alpha _{me}$ the magnetoelectric coefficient.
This equation implies an
enhancement of the bare relativistic effect in the condensed matter, i.e., it is due to
the newly proposed magnetoelectric effect.\cite{katsura05}
It should be noted that the suitable materials for the observation of polarization
are magnetic insulators and semiconductors because the induced electric polarization
in a good metal would be completely screened by free electrons.

In the cases of the magnetic ring with the tangent magnetization along the
ring or the constant radial magnetization, using (21) we find
that the polarization vector ${\bf P}$ is oriented along the radius of the
ring, and $P_r=-\alpha _{me}/R$, where $R$ is the ring radius.

Let us estimate the magnitude of effect for a typical magnetic
nanostructure, namely, a ferromagnetic disc of radius $R$ and
thickness $h$ with a circular vortex domain (tangential
magnetization).\cite{shinjo00} The effect of the vortex can be
approximated by replacing the disc by a ring of internal radius
$r$ equal to the radius of the vortex core. In this case the
polarization ${\bf P}$ is radial and decays as the inverse
distance from the ring axis. We can approximately estimate the
electrostatic potential $U$ at a point located on the axis at an
altitude $z$ above the disc in the limit of $r,h\ll |z|\ll R$. We
obtain $U\simeq -\alpha _{me}h/|z|$. We estimate $\alpha _{me}$
using the result of a three-atom model from
Ref.~[\onlinecite{katsura05}] for the polarization per unit
volume, which gives $\alpha _{me}\simeq (e/a)\, (V/\Delta )^3$.
Here $a$ is the lattice constant, $V$ is a hopping energy (hopping
between the transition and oxygen atoms in
[\onlinecite{katsura05}]), and $\Delta $ is a characteristic
electron energy (between $d$ and $p$ orbitals of different atoms).
This formula was obtained for large SO splitting $\Delta _{SO}\gg
\Delta $, and it does not explicitly depend the SO interaction.
For a magnetic disc, using typical values $R=1\, \mu$m, $h=10$~nm,
and $r=10$~nm, this yields $U\simeq 0.1\, (V/\Delta )^3$~V at an
altitude $|z|=100$~nm. The vortex can be removed by application of
an external magnetic field. The effect can be possibly detected
experimentally by using a single-electron transistor as a detector
of electric field.

We calculated the equilibrium spin current in mesoscopic
systems, and showed that it does not vanish in
the metastable state of magnetic system with topologically nontrivial
magnetization profile. The electric polarization induced by the
spin current can be experimentally measured.

This work is supported by FCT Grant POCI/FIS/58746/2004
(Portugal) and by Polish Grants PBZ/KBN/044/P03/2001 and 2~P03B~053~25.
V.D. thanks the Calouste Gulbenkian Foundation
in Portugal for support.


\begin{thebibliography}{99}

\bibitem{prinz98}
G.A. Prinz, Science {\bf 282}, 1660 (1998); S.A. Wolf {\it et
al.}, Science {\bf 294}, 1488 (2001); I. $\check{\rm Z}$uti\'c, J.
Fabian, S. Das Sarma, Rev. Mod. Phys. {\bf 76}, 323 (2004).

\bibitem{ohno99}
Y. Ohno {\it et al.}, Nature {\bf 402}, 790 (1999).

\bibitem{ganichev02}
S.D. Ganichev {\it et al.}, Nature {\bf 417}, 153 (2002).

\bibitem{konig01}
J. K\"onig {\it et al.}, Phys. Rev. Lett. {\bf 87}, 187202 (2001).

\bibitem{hirsch99}
J.E. Hirsch, Phys. Rev. Lett. {\bf 83}, 1834 (1999); S. Murakami,
N. Nagaosa, and S.C. Zhang, Science {\bf 301}, 1348 (2003); J.
Sinova {\it et al.}, Phys. Rev. Lett. {\bf 92}, 126603 (2004).

\bibitem{pareek04}
T.P. Pareek, Phys. Rev. Lett. {\bf 92}, 076601 (2004).

\bibitem{meier03}
F. Meier and D. Loss, Phys. Rev. Lett. {\bf 90}, 167204 (2003).

\bibitem{schutz03}
F. Sch\"utz, M. Kollar, and P. Kopietz, Phys. Rev. Lett. {\bf 91},
017205 (2003); Phys. Rev. B {\bf 69}, 035313 (2004).

\bibitem{katsura05}
H. Katsura, N. Nagaosa, and A.V. Balatsky, Phys. Rev. Lett. {\bf
95}, 057205 (2005).

\bibitem{zhang05}
P. Zhang {\it et al}, cond-mat/0503305 (2005).

\bibitem{rashba03}
E.I. Rashba, Phys. Rev. B {\bf 68}, 241315(R) (2003); {\bf 70},
161201(R) (2004); cond-mat/0408119; cond-mat/0507007.

\bibitem{schutz04}
F. Sch\"utz, P. Kopietz, and M. Kollar,
Eur. Phys. J. B {\bf 41}, 557 (2004).

\bibitem{chandra90}
P. Chandra, P. Coleman, and A. I. Larkin, J. Phys. Cond. Mat. {\bf
2}, 7933 (1990).

\bibitem{bruno04}
P. Bruno, Phys. Rev. Lett. {\bf 93}, 247202 (2004); Erratum-{\it
ibid.} {\bf 94}, 239903 (2005); V.K. Dugaev, P. Bruno, B. Canals,
and C. Lacroix, Phys. Rev. B {\bf 72}, 024456 (2005).

\bibitem{itzykson}
C. Itzykson and J.-B. Zuber,
{\em Quantum Field Theory} (McGraw-Hill, New York, 1980).

\bibitem{pokrovsky88}
V.L. Pokrovsky, M.V. Feigel'man, and A.M. Tsvelick, In: {\em Spin
Waves and Magnetic Excitations 2}, edited by A.~S.~Borovik-Romanov
and S.~K.~Sinha (Elsevier, Amsterdam, 1988), p.~67.

\bibitem{durst00}
A.C. Durst and P.A. Lee, Phys. Rev. B {\bf 62}, 1270 (2000).

\bibitem{tatara03}
G. Tatara and H. Kohno, Phys. Rev. B {\bf 67}, 113316 (2003);
G. Tatara and N. Garcia, Phys. Rev. Lett. {\bf 91}, 076806 (2003).

\bibitem{wang04}
B. Wang {\it et al.}, Phys. Rev. B {\bf 69}, 174403 (2004).

\bibitem{frenkel26}
J. Frenkel, J. Physik {\bf 37}, 243 (1926).

\bibitem{shinjo00}
T. Shinjo {\it et al.}, Science {\bf 289}, 930 (2000);
A. Wachowiak {\it et al.}, Science {\bf 298}, 577 (2004).

\end{thebibliography}
\end{document}